\documentclass[preprint,showpacs,preprintnumbers,amsmath,amssymb,amssymb]{revtex4}

\usepackage{graphicx}
\usepackage{dcolumn}
\usepackage{bm}

\begin{document}

\hyphenpenalty=5000
\tolerance=1000
\makeatletter


\title{Search for the Higgs Boson $H_2^0$ at LHC in 3-3-1 Model}   
\author{J.\ E.\ Cieza Montalvo$^1$}   
\affiliation{$^1$Instituto de F\'{\i}sica, Universidade do Estado do Rio de Janeiro, Rua S\~ao Francisco Xavier 524, 20559-900 Rio de Janeiro, RJ, Brazil}
\author{R. J. Gil Ram\'{i}rez, G. H. Ram\'{i}rez Ulloa, A. I. Rivasplata Mendoza$^2$}   
\affiliation{$^2$Universidad Nacional de Trujillo, Departamento de F\'{i}sica, Av. Juan Pablo II S/N; Ciudad Universitaria, Trujillo, La Libertad, Per\'u}  
\author{M. D. Tonasse$^{3}$\footnote{Permanent address: Universidade Estadual Paulista, {\it Campus} Experimental de Registro, Rua Nelson Brihi Badur 430, 11900-000 Registro, SP, Brazil}}   
\affiliation{$^3$Instituto de F\'\i sica Te\'orica, Universidade Estadual Paulista, \\  Rua Dr. Bento Teobaldo Ferraz 271, 01140-070 S\~ao Paulo, SP, Brazil}   
\date{\today}
   
\begin{abstract}
We present an analysis of production and signature of neutral Higgs boson ($H_{2}^{0}$) on the version of the 3-3-1 model containing heavy leptons at the Large Hadron Collider. We studied the possibility to identify it using the respective branching ratios. Cross section are given for the collider energy, $\sqrt{s} =$ 14 TeV. Event rates and significances are discussed for two possible values of integrated luminosity, 300 fb$^{-1}$ and 3000 fb$^{-1}$.

\end{abstract}
\maketitle

\section{INTRODUCTION \label{introd}}

The way to the understanding of the symmetry breaking mainly go 
through the scalars, although there are many other models that not contain elementary scalar fields, such as Nambu-Jona-Lasinio mechanism, technicolour theories, the strongly interacting gauge systems  \cite{nambu}. These  scalars protect the renormalizability of the theory by moderating the cross section growth. But so far, despite many experimental and theoretical efforts in order to understanding the scalar sector, the Higgs mechanism remains still unintelligible.  Nowadays, the major goal of the experimentalists in particle physics at the Large Hadron Collider (LHC), is to unravel the nature of electroweak symmetry breaking. The Standard Model (SM) is the prototype of a gauge theory with spontaneous symmetry breaking. This had great success in explaining the most of the experimental data.  However, recent results from neutrino osclillation experiments makes clear that the SM is not complete, then the neutrino oscillation implies that at least two neutrino flavors are massive. Moreover, there are others crucial problems in particle physics that not get response in SM. For instance, it offers not solution to the dark matter problem, dark energy and the asymmetry of matter-antimatter in the Universe. 
Therefore, there is a consensus among the particle physicists that the SM must be extended. \par 

In the SM appear only one elementary scalar,  which arises through the breaking of electroweak symmetry and this is the Higgs boson. The Higgs Boson is an important prediction of several quantum field theories and is so crucial to our understanding of the Universe. So on 4 July 2012, a previously unknown particle with a mass around 126 GeV was announced as being detected, which physicists suspected at the time to be the Higgs boson \cite{atlas1, atlas11, cms}. By March 2013, the particle has been proven to be the Higgs boson, because it behave, interact and decay in the ways predicted by the SM, and was also tentatively confirmed to have positive parity and zero spin, two fundamental criteria of a Higgs boson, making it also the first known scalar particle to be discovered in nature. \par 

Different types of Higgs bosons, if they exist, may lead us into new realms of physics beyond the SM. Since the SM leaves many questions open, there are several  extensions. For example, if the Grand Unified Theory (GUT) contains the SM at high energies, then the Higgs bosons associated with GUT symmetry breaking must have masses of order $M_{X} \sim {\cal O} (10^{15})$
GeV. Supersymmetry \cite{supers} provides a solution to hierarchy problem through the cancellation of the quadratic divergences via fermionic and bosonic loops contributions \cite{cancell}. Moreover, the Minimal Supersymmetric extension of the SM can be derived as an effective theory of supersymmetric GUT \cite{sgut}. \par 

Among these extensions of the SM there are also other class of models based on SU(3)$_C \otimes$SU(3)$_L \otimes$U(1)$_N$ gauge symmetry (3-3-1 model) \cite{PT93a, PP92, FR92}, where the anomaly cancellation mechanisms occur when the three basic fermion families are considered and not family by family as in the SM. This mechanism is peculiar because it requires that the number of families is an integer multiple of the number of colors. This feature combined together  with the asymptotic freedom, which is a property of quantum chromodynamics, requires that the number of families is three. Moreover, according to these models, the Weinberg angle is restricted to the value $s_W^2 = \sin^2\theta_W <1/4$ in the version of heavy-leptons \cite {PT93a}. Thus, when it evolves to higher values, it shows that the model loses its perturbative character when it reaches to mass scale of about 4 TeV \cite{DI05}. Hence, the 3-3-1 model is one of the most interesting extensions of the SM and is phenomenologically well motivated to be probed at the LHC and other accelerators. \par 

In this work we study the production and signatures of an extended neutral Higgs boson $H_{2}^{0}$, predicted by the 3-3-1 model, which incorporates the charged heavy leptons \cite{PT93a, TO96}. We can show that the neutral Higgs boson signatures can be significant at LHC. The signal of the new particle can be obtained by studying the different decay modes and if we consider a luminosity of 10 times higher than original LHC design. With respect to both mechanisms, that is the Drell-Yan and gluon-gluon fusion, we consider the $Z^\prime$, $H_1^0$ and $H_2^0$ as propagators. Therefore in Sec. II we present the relevant features of the model. In Sec. III we compute the total cross sections of the process and in Sec. IV we summarize our results and conclusions.


\section{Relevant Features of the Model \label{sec2}}

We are working here with the version of the 3-3-1 model that contains heavy leptons \cite{PT93a}. The model is based on the semi simple symmetry group SU(3)$_C$$\otimes$SU(2)$_L$$\otimes$U(1)$_N$. The electric charge operator is given by

\begin{equation}
\frac{Q}{e} = \left(T_3 - \sqrt{3} \ T_8\right) + N,
\label{op}\end{equation}
where $T_3$ and $T_8$ are the generators of SU(3) and $e$ is the elementary electric charge. So, we can build three triplets of quarks of SU(3)$_L$ such that 
\begin{equation}
Q_{1L} = \left(\begin{matrix} u^\prime_1 \\ d^\prime_1 \\ J_1\end{matrix}\right)_L \sim \left({\bf 3}, \frac{2}{3}\right), \qquad Q_{\alpha L} = \left(\begin{matrix} J^\prime_\alpha \\ u^\prime_\alpha \\ d^\prime_\alpha\end{matrix}\right)_L \sim \left({\bf 3}^*, -\frac{1}{3}\right).
\label{quarks}\end{equation}
where the new quark $J_1$ carries $5/3$ units of electric charge while $J_\alpha$ $\left(\alpha = 2, 3\right)$ carry $-4/3$ each. We must also introduce the right-handed fermionic fields $U_R \sim \left({\bf 1}, 2/3 \right)$, $D_R \sim \left({\bf 1}, -1/3 \right)$, $J_{1R} \sim \left({\bf 1}, 5/3 \right)$ and $J^\prime_{\alpha R} \sim \left({\bf 1}, -4/3 \right)$. We have defined $U = \left(\begin{matrix}u^\prime & c^\prime & t^\prime\end{matrix}\right)$ and $D = \left(\begin{matrix}d^\prime & s^\prime & b^\prime\end{matrix}\right)$. \par

The spontaneous symmetry breaking is accomplished {\it via} three SU(3) scalar triplets, which are,
 
\begin{equation}
\eta = \left(\begin{matrix} \eta^0 \\ \eta^-_1 \\ \eta^+_2\end{matrix}\right) \sim \left({\bf 3}, 0\right), \quad \rho = \left(\begin{matrix} \rho^+ \\ \rho^0 \\ \rho^{++}\end{matrix}\right) \sim \left({\bf 3}, 1\right), \quad \chi = \left(\begin{matrix} \chi^- \\ \chi^{--} \\ \chi^0\end{matrix}\right) \sim \left({\bf 3}, -1\right).
\label{rc}\end{equation}

For sake of simplicity, we will assume here that the model respects the $B + L$ symmetry, where $B$ is the baryon number and $L$ is the lepton number. Then, the more general renormalizable Higgs potential is given by
\begin{eqnarray}
V\left(\eta, \rho, \chi\right) & = & \mu_1^2\eta^\dagger\eta + \mu_2^2\rho^\dagger\rho + \mu_3^2\chi^\dagger\chi +  \lambda_1\left(\eta^\dagger\eta\right)^2 + \lambda_2\left(\rho^\dagger\rho\right)^2 + \lambda_3\left(\chi^\dagger\chi\right)^2 + \cr 
&& + \eta^\dagger\eta\left[\lambda_4\left(\rho^\dagger\rho\right) + \lambda_5\left(\chi^\dagger\chi\right)\right] + \lambda_6\left(\rho^\dagger\rho\right)\left(\chi^\dagger\chi\right) + \lambda_7\left(\rho^\dagger\eta\right)\left(\eta^\dagger\rho\right) + \cr 
&& + \lambda_8\left(\chi^\dagger\eta\right)\left(\eta^\dagger\chi\right) + \lambda_9\left(\rho^\dagger\chi\right)\left(\chi^\dagger\rho\right) + \frac{1}{2}\left(f\varepsilon^{ijk}\eta_i\rho_j\chi_k + {\mbox{c. H.}}\right),
\label{pot}\end{eqnarray}
where $\mu_i$ $\left(i = 1, 2, 3\right)$ and $f$ are constants with mass dimension and $\lambda_j$ $\left(j = 1, \ldots, 9\right)$ are dimensionless constants \cite{TO96}. The potential (\ref{pot}) is bounded from below when the neutral Higgs fields develops the vacuum  expectation values (VEVs) $\langle\eta^0\rangle = v_\eta$, $\langle\rho^0\rangle = v_\rho$ and $\langle\chi^0\rangle = v_\chi$, with $v_\eta^2 + v_\rho^2 = v_W^2 = 246^2$ GeV$^2$. The scalar $\chi^0$ is supposedly heavy and it is responsible for the   spontaneous symmetry breaking of SU(3)$_L$$\otimes$U(1)$_N$ to SU(2)$_L$$\otimes$U(1)$_Y$ of the standard model. Meanwhile, $\eta^0$ and $\rho^0$ are lightweight and are responsible for the breaking of SU(2)$_L$$\otimes$U(1)$_Y$ to U(1)$_Q$, of the electromagnetism. Therefore, it is reasonable to expect 
\begin{equation}
v_\chi \gg v_\eta, v_\rho.
\label{ap}\end{equation}
The potential (\ref{pot}) provides the masses of neutral Higgs as
\begin{subequations}\begin{align}
m^2_{H_1^0} & \approx 4\frac{\lambda_2 v_\rho^4 - \lambda_1 v_\eta^4}{v_\eta^2 - v_\rho^2}, & m^2_{H_2^0} &\approx \frac{v_W^2v_\chi^2}{2v_\eta v_\rho}, \\
m^2_{H_3^0} & \approx -\lambda_3 v_\chi, & m^2_h & = -\frac{fv_\chi}{v_\eta v_\rho}\left[v_W^2 + \left(\frac{v_\eta v_\rho}{v_\chi}\right)^2\right]
\end{align}\label{mas}\end{subequations}

with the corresponding eigenstates
\begin{equation}
\left(\begin{matrix} \xi_\eta \\ \xi_\rho\end{matrix}\right) \approx \left(\begin{matrix} c_w & s_w \\ s_w & c_w\end{matrix}\right)\left(\begin{matrix} H_1^0 \\ H_2^0 \end{matrix}\right), \quad \xi_\chi \approx H_3^0, \quad \zeta_\chi \approx ih,
\label{auto}\end{equation}
where the mixing parameters are $c_w = \cos w = v_\eta/\sqrt{v_\eta^2 + v_\rho^2}$ and $s_w = \sin w$ \cite{TO96}. In Eqs. (\ref{mas}) and (\ref{auto}) we used the approximation (\ref{ap}) and, for not to introduce the new mass scale in the model, we assume $f \approx -v_\chi$. We can then note that $H^0_3$ is a typical 3-3-1 Higgs boson. The scalar $H^0_1$ is one that can be identified with the SM Higgs, since its mass and eigenstate do not depend on $v_\chi$. \par
Now, we can write the Yukawa interactions for the ordinary quarks, i.e.
\begin{eqnarray}
{\cal L}^Y_{q} & = & \sum_\alpha\left[\overline{Q}_{1L}\left(G_{1\alpha}U^\prime_{\alpha R}\eta + \tilde{G}_{1\alpha}D^\prime_{\alpha R}\rho\right) + \sum_i\overline{Q}_{iL}\left(F_{i\alpha}U^\prime_{\alpha R}\rho^* + \tilde{F}_{i\alpha}D^\prime_{\alpha R}\eta^*\right)\right],
\label{Lq}
\end{eqnarray}
where $G_{ab}$ and $G^\prime_{ab}$ ($a$ and $b$ are generation indexes) are coupling constants. \par
The interaction eigenstates (\ref{quarks}) and their right-handed counterparts can rotate about their respective physical eigenstates as 
\begin{subequations}\begin{eqnarray}
&& U^\prime_{aL\left(R\right)} = {\cal U}_{ab}^{L\left(R\right)}U_{bL\left(R\right)}, \\
&& D^\prime_{aL\left(R\right)} = {\cal D}_{ab}^{L\left(R\right)}U_{bL\left(R\right)}, \quad J^\prime_{aL\left(R\right)} = {\cal J}_{ab}^{L\left(R\right)}J_{bL\left(R\right)} \ .
\end{eqnarray}\end{subequations}

Since the cross sections involving the sum over the flavors and rotation matrices are unitary, then the mixing parameters have no major effect on the calculations. In terms of physical fields the Yukawa Lagrangian for the neutral Higgs can be written as

\begin{eqnarray}
-{\cal L}_Q & = & \frac{1}{2}\left\{\overline{U}\left(1 + \gamma_5\right)\left[1 + \left[\frac{s_w}{v_\rho} + \left(\frac{c_w}{v_\eta} + \frac{s_w}{v_\rho}\right){\cal V}^U\right]H_1^0 + \right.\right. \cr
&& \left.\left. + \left[\frac{-c_w}{v_\rho} + \left(\frac{s_w}{v_\eta} - \frac{c_w}{v_\rho}\right){\cal V}^U\right]H_2^0\right]M^UU + \right. \cr
&& \left. + \overline{D}\left(1 + \gamma_5\right)\left[1 + \left[\frac{c_w}{v_\eta} + \left(\frac{s_w}{v_\rho} - \frac{c_w}{v_\eta}\right){\cal V}^D\right]H_1^0 + \right.\right. \cr
&& \left.\left. + \left[\frac{s_w}{v_\eta} - \left(\frac{c_w}{v_\rho} + \frac{s_w}{v_\eta}\right){\cal V}^D\right]H_2^0\right] M^DD\right\} + {\mbox{H. c.}},
\end{eqnarray}\label{Yuk}
where $V^U_LV^D_L = V_{\rm CKM}$ is the Cabibbo-Kobayashi-Maskawa matrix, ${\cal V}^U$ and ${\cal V}^D$ are arbitrary mixing matrices and $M^U = {\rm diag}\left(\begin{matrix} m_u & m_c & m_t\end{matrix}\right)$ and $M^D = {\rm diag}\left(\begin{matrix} m_d & m_s & m_b\end{matrix}\right)$ are matrices which carrying the masses of the quarks. \par

In the gauge sector, beyond the standard particles $\gamma$, $Z$, and $W^\pm$ the model predicts: one neutral $\left(Z^\prime\right)$, two single-charged $\left(V^\pm\right)$, and two doubly-charged $\left(U^{\pm\pm}\right)$ gauge bosons. The gauge interactions with  Higgs  bosons are given by 
\begin{equation}
{\cal L}_{GH} = \sum_\varphi\left({\cal D_\mu\varphi}\right)^\dagger\left({\cal D_\mu\varphi}\right),
\label{deri}\end{equation}
where the covariant derivatives are
\begin{equation}
{\cal D}_\mu\varphi_i = \partial_\mu\varphi_i - ig\left({W}_\mu.\frac{{ T}}{2}\right)^j_i\varphi_j - ig^\prime N_\varphi\varphi_iB_\mu,
\end{equation}
where $\varphi = \eta$, $\rho$, $\chi$ $\left(N_\eta = 0, N_\rho = 1, N_\chi = -1\right)$ are the Higgs triplets, ${W}_\mu$ and $B_\mu$ are the SU(2) and U(1) field tensors, $g$ and $g^\prime$ are the U(1) and SU(2) coupling constants, respectively. Diagonalization of the Lagrangean (\ref{deri}), after symmetry 
breaking, gives masses for the neutral weak gauge bosons, i.e.,

\begin{equation}
m_Z \approx \frac{\vert e\vert}{2s_Wc_W}v_W,     \qquad m_{Z^\prime}^2 \approx \frac{1}{3\left(1 - 4s_W^2\right)}\left(\frac{\vert e\vert c_Wv_\chi}{s_W}\right)^2,
\label{masszz}\end{equation}

where $s_W = \sin{\theta_W}$, with $\theta_W$ being the Weinberg angle, and $c_W^2 = 1 - s_W^2$. Then the eigenstates are
\begin{subequations}\begin{eqnarray}
W^3_\mu & \approx & s_WA_\mu - c_WZ_\mu \\
W^8_\mu & \approx & -\sqrt{3}s_W\left(A_\mu - \frac{s_W}{c_W}Z_\mu\right) + \frac{\sqrt{1 - 4s_W^2}}{c_W}Z^\prime_\mu \\
B_\mu & \approx & \frac{s_W}{\sqrt{1 - 4s_W^2}}A_\mu + \frac{s_W}{c_W}\left(Z_\mu + \sqrt{3}Z^\prime_\mu\right).
\end{eqnarray}\label{eig}\end{subequations}
In Eqs. (\ref{masszz}) and (\ref{eig}) we have used the approximation (\ref{ap}). Finally the weak neutral current in the sector of u and d quarks reads 

\begin{subequations}\begin{eqnarray}
-{\cal L}_Z & = & \frac{\left|e\right|}{2s_Wc_W}\overline{q}\gamma^\mu\left[v\left(q\right) + a\left(q\right)\gamma_5\right]qZ_\mu \\
-{\cal L}_{Z^\prime} & = & \frac{\left|e\right|}{2s_Wc_W}\overline{q}\gamma^\mu\left[v^\prime\left(q\right) + a^\prime\left(q\right)\gamma_5\right]qZ^\prime_\mu
\end{eqnarray}\label{ZZl}\end{subequations}
whose coefficients are
\begin{subequations}\begin{align}
v\left(u\right) & = 1 - \frac{s_W^2}{8}, & a\left(u\right) & = -a\left(d\right) = -1, & v\left(d\right) & = -1 + \frac{4}{3}s_W^2, \\
v^\prime\left(u\right) & = \sqrt{1 + 4s_W^2}, & a^\prime\left(u\right) & = \sqrt{\frac{1 - 4s_W^2}{3}},  & v^\prime\left(d\right) & = \frac{2s_W^2 - 1}{\sqrt{3}}, \\
a^\prime\left(d\right) & = -v^\prime\left(d\right).
\end{align}\label{coefq}\end{subequations}

In this work we study the production of the neutral Higgs boson $H_{2}^{0}$ at $pp$ colliders. With respect to both mechanisms, that is the Drell-Yan and gluon-gluon fusion, we consider the $Z^\prime$, $H_1^0$ and $H_2^0$ as propagators.

\section{CROSS SECTION PRODUCTION}   

The mechanisms for the production of a neutral Higgs particle $H_{2}^{0}$ in $pp$ collisions occurs in association with the boson $Z^{\prime}$, $H_{1}^{0}$ and $H_{2}^{0}$, see Fig. 1 and Fig. 2. Unlike of the SM, where the gluon-gluon fusion dominates over Drell–Yan when the Higgs boson is heavier than $100$ GeV \cite{barger}, in 3-3-1 Model does  not occur. Here the mechanism of Drell-Yan dominates over gluon-gluon fusion at leading order for $H_{2}^{0}$ production at $\sqrt{s}=14$ TeV. The process $pp \to  H_{2}^{0} Z$ $(i = 1, 2)$ takes place in the $s$ channel. The term involving the boson $Z$ is absent, because there is no coupling between the $Z$ and $H_{2}^{0} Z$, moreover the interference term between the $Z^{\prime}$ and $H_{2}^{0}$ should be absent because it gives an imaginary value. So using the interaction Lagrangian \cite{PT93a, cieton2}, we obtain the differential cross section in the first place for Drell-Yan for $H_{2}^{0}$

\begin{eqnarray} 
\frac{d \hat{\sigma}_{H_{2}^{0}}}{d\cos \theta} & = &\frac{\beta_{H_{2}^{0}} \ g^{2}}{192 \pi c_{W}^{2} s} \Biggl \{ \frac{g^4 \ \Lambda_{ ZZ'H_{2}^{0}}^{2}}{4(s- m_{Z'}^{2}+ im_{Z'} \Gamma_{Z'})^{2}}
\left ((m_{Z}^{2}+ \frac{tu}{m_{Z}^{2}}- t- u + s)(g_{{V'}^{q}}^{2}+ g_{{A'}^{q}}^{2}) \right )   \nonumber  \\ 
&& + \left( \frac{v_{\eta}^{2} \ v_{\rho}^{2} m_{q}^{2} }{2 \ v_{W}^{6}}  |\chi^{(1)}(\hat{s})|^{2} + \frac{\left (m_{u} \frac{v_{\eta}}{v_{\rho}}- m_{d} \frac{v_{\rho}}{v_{\eta}} \right )^{2} (v_{\rho}^{2}- v_{\eta}^{2})^2}{32 \ v_{W}^{6}} |\chi^{(2)}(\hat{s})|^{2} \right.  \nonumber  \\
&&\left.  + \frac{m_{q}  \left (m_{u} \frac{v_{\eta}}{v_{\rho}}- m_{d} \frac{v_{\rho}}{v_{\eta}} \right ) \ v_{\eta} v_{\rho} (v_{\rho}^{2}-v_{\eta}^{2})}{4 \ v_{W}^{6}} |\chi^{(1)}(\hat{s})|  |\chi^{(2)}(\hat{s})| \right)  \nonumber \\   
&& \left(\frac{\hat{s}}{m_{Z}^{2}} (\hat{s}^{2}-2 m_{Z}^{2} \ \hat{s}+  m_{Z}^{4}) - \frac{m_{q}^{2}}{m_{Z}^{2}} (2 \hat{s}^{2} - 4 \hat{s} \ m_{Z}^{2} + 2  m_{Z}^{4} )  \right. +  \nonumber \\
&&\left. -\frac{m_{H_{2}^{0}}^{2}}{m_{Z}^{2}} (2 \hat{s}^{2} + 2 \hat{s} m_{q}^{2}- 4 m_{q}^{2} \hat{s} - 4 m_{q}^{2} m_{Z}^{2} ) + \frac{m_{H_{2}^{0}}^{4} \hat{s}}{m_{Z}^{2}}- \frac{2 m_{H_{2}^{0}}^{4} m_{q}^{2}}{m_{Z}^{2}}     \  \right)  \Biggr \}  , 
\label{DZZ'H2}
\end{eqnarray} 
here g is the coupling constant of the weak interaction, the $\beta_{H_{2}^{0}}$ is the Higgs velocity in the c.m. of the subprocess which is equal to
\[
\beta_{H_{2}^{0}} = \frac{ \left [\left( 1- \frac{(m_{Z}+ m_{H_{2}^{0}})^{2}}{\hat{s}} \right) \left(1- \frac{(m_{Z}- m_{H_{2}^{0}})^{2}}{\hat{s}} \right) \right ]^{1/2}}{1-\frac{m_{Z}^{2}-m_{H_{2}^{0}}^{2}}{\hat{s}}}  \ \ ,
\]

and $t$ and $u$ are

\[
t  = m_{q}^{2}+ m_{Z}^{2} - \frac{s}{2} \Biggl \{ \left(1+ \frac{m_{Z}^{2}- m_{H}^{2}}{s}\right)- \cos \theta  \left [\left( 1- \frac{(m_{Z}+ m_{H})^{2}}{s} \right) \left(1- \frac{(m_{Z}- m_{H})^{2}}{s} \right) \right ]^{1/2}\Biggr \}, 
\]

\[  
u  = m_{q}^{2}+ m_{H}^{2} - \frac{s}{2} \Biggl \{ \left(1- \frac{m_{Z}^{2}- m_{H}^{2}}{s}\right)+ \cos \theta  \left [\left( 1- \frac{(m_{Z}+ m_{H})^{2}}{s} \right) \left(1- \frac{(m_{Z}- m_{H})^{2}}{s} \right) \right ]^{1/2}\Biggr \}, 
\]
where $\theta$ is the angle between the Higgs and the incident quark in the CM frame. \par 

And we have also defined

\[
\chi^{i}(\hat{s}) = \frac{1}{\hat{s}-m_{H_{i}^{0}}^{2} + i m_{H_{i}^{0}} \Gamma_{H_{i}^{0}}}  \ ,
\]
with $\Gamma_{H_{i}^{0}}$ being the Higgs boson total width, $i=1,2$, $\Gamma_{Z'}$ \cite{cieton1,cieton2}, are the total width of the $Z'$ boson, $m_{q}$, were $q=u,d$ are the masses of the quark, $g_{V', A'}^{q}$ are the 3-3-1 quark coupling constants, $\sqrt{\hat{s}}$ is the center of mass energy of the  $q \bar{q}$ system, $g= \sqrt{4 \ \pi \ \alpha}/\sin \theta_{W}$ and $\alpha$ is the fine structure constant, which we take equal to $\alpha=1/128$. For the $Z^\prime$ boson we take  $M_{Z^\prime} = \left(0.5 - 3\right)$ TeV, since $M_{Z^\prime}$ is proportional to the VEV $v_\chi$ \cite{PP92,FR92}. For the standard model parameters, we assume Particle Data Group values, {\it i. e.}, $M_Z = 91.19$ GeV, $\sin^2{\theta_W} = 0.2315$, and $M_W = 80.33$ GeV 
\cite{Nea10}, $\it{t}$ and $\it{u}$ are the kinematic invariants. We have also defined the $\Lambda_{ZZ^{\prime} H_{2}^{0}}$ as the coupling constants of the $Z^{\prime}$ boson to Z boson and Higgs $H_{2}^{0}$, the $\Lambda_{H_{i}^{0}H_{2}^{0}Z}$ are the couplings constants of the $H_{1}^{0}$ boson to $H_{2}^{0}$ and Z boson and of the  $H_{2}^{0}$ boson to $H_{2}^{0}$ and Z boson, these coupling constants should be multiplied by $p^{\mu}-q^{\mu}$ to get a $\Lambda_{H_{i}^{0}H_{2}^{0}Z}^{\mu} = \Lambda_{H_{i}^{0}H_{2}^{0}Z} (p^{\mu}-q^{\mu})$ with $p$  and $q$ being the momentum four-vectors of the $H_{2}$ and $Z$ boson, and the $\Lambda_{q \bar{q} H_{i}^{0}}$ are the coupling constants of the $H_{1}^{0}(H_{2}^{0})$ to $q \bar{q}$, the $v\left(q\right)$, $a\left(q\right)$, $v^\prime\left(q\right)$ and $a^\prime\left(q\right)$ are given in \cite{cieton1}. We remark still that in 3-3-1 model, the states $H_1^0$ and $H_2^0$ are mixed. \par

\begin{subequations}\begin{eqnarray}
\left(\Lambda_{q\bar{q}Z^\prime}\right)_\mu & \approx & i\frac{\vert e\vert}{2s_Wc_W}\gamma_\mu\left[v^\prime\left(q\right) + a^\prime\left(q\right)\gamma_5\right], \\
\Lambda_{q\bar{q}H_1^0} & \approx & -i\frac{m_q}{2v_W}\left(1 + \gamma_5\right), \\
\Lambda_{q\bar{q}H_2^0} & \approx & \frac{i}{2v_W}\left(-m_u\frac{v_\eta}{2v_\rho} + m_d\frac{v_\rho}{v_\eta}\right)\left(1 + \gamma_5\right) \\
\left(\Lambda_{ZZ^\prime H_2^0}\right)_{\mu\nu} & \approx & \frac{g^2}{\sqrt{3}\left(1 - 4s_W^2\right)}\frac{v_\eta v_\rho}{v_W}g_{\mu\nu}, 
\\
\left(\Lambda_{H_2^0H_2^0Z}\right)_\mu & \approx & -\frac{g}{2}\frac{m_Z}{m_W}\frac{\left(v_\rho^2 - v_\eta^2\right)}{v_W^2}\left(p - q\right)_\mu, \\ 
\left(\Lambda_{H_1^0H_2^0Z}\right)_\mu & \approx & -2g\frac{m_Z}{m_W}\frac{v_\rho v_\eta}{v_W^2}\left(p - q\right)_\mu,
\label{eigc}\end{eqnarray}\label{eigthen}\end{subequations}

The total cross section for the process $pp \rightarrow qq \rightarrow Z H_{2}^{0}$ is related to the subprocess $qq \rightarrow Z H_{2}^{0}$ total cross  section $\hat{\sigma}$, through   
\begin{equation}    
\sigma = \int_{\tau_{min}}^{1} \int_{\ln{\sqrt{\tau_{min}}}}^{-\ln{\sqrt{\tau_{min}}}} d \ \tau \ dy \   q\left(\sqrt{\tau}e^y, Q^2\right) q\left(\sqrt{\tau}e^{-y}, Q^2\right)   \hat{\sigma}\left(\tau, s\right),   
\end{equation}\noindent
where $\tau_{min} = (m_{Z}+ m_{H_{2}^{0}})^{2}/s (\tau =\hat{s}/s )$ and  $q\left(x,Q^2\right)$ is the 
quark structure function.\par 

Another form to produce a neutral Higgs is {\it via} the gluon-gluon  fusion, namely through the  reaction of the type $pp \rightarrow g g   \rightarrow Z H_{2}^{0}$. Since the final state is neutral, the $s$ channel involves  the exchange of the boson $Z'$, $H_{1}^{0}$ and $H_{2}^{0}$. The exchange of a photon is not allowed by C conservation (Furry's theorem), which also indicates that only the axial-vector coupling of the boson $Z'$, contribute to this process. It is important to emphasis that for production of $H_{2}^{0}$ we take the interference between the $Z^{\prime}$, which are antisymmetric in the gluon polarizations and the $H_{2}^{0}$, we only consider the antisymmetric term of $H_{2}^{0}$, because the other part is symmetric and therefore vanishes, then we write explicity the $Z^{\prime}, H_{1}^{0}$ and $H_{2}^{0}$ contributions to the elementary cross section for production of $H_{2}^{0}$

\begin{eqnarray} 
\left (\frac{d \hat{\sigma}}{d\cos \theta} \right )_{pp \rightarrow Z H_{2}^{0}}^{Z'} & = & \frac{g^{6} \ \alpha_s^2  \ (\Lambda_{Z(Z^{\prime})ZH_{i}^{0}})^{2} \ \Delta}{8192 \ \pi^{3}  \hat{s} \  c_W^{2}  M_{Z(Z')}^4} \beta_{H_{i}^{0}} \left|
\sum_{q=u,d} T_3^q(q') \left( 1 + 2 \delta_q I_q  \right) \right|^2   ,
\end{eqnarray}

\begin{eqnarray} 
\left (\frac{d \hat{\sigma}}{d\cos \theta} \right )_{pp \rightarrow Z H_{2}^{0}}^{H_{1}^{0}-H_{2}^{0}} & = &  \frac{g^{2} \ \alpha_s^2 \ (v_{\rho}^{2} - v_{\eta}^{2}) v_{\rho} v_{\eta} \ \Omega \ \beta_{H_{i}^{0}}}{8192 \ \pi^{3} \ \hat{s} \  c_W^{2} v_{W}^{6}} {\it Re}  \  \chi^{(1)}(\hat{s}) \ \chi^{(2)}(\hat{s})     \nonumber  \\
&& \sum_{q=u,d}  m_{q}^{3} \left (m_{u} \frac{v_{\eta}}{v_{\rho}}- m_{d} \frac{v_{\rho}}{v_{\eta}} \right )   I_q  \ \sum_{q=u,d}  I_q^{*}   \nonumber \ +  \ \frac{g^{2} \ \alpha_s^2 \ (v_{\rho}^{2} - v_{\eta}^{2}) v_{\rho} v_{\eta} \ \hat{s} \ \Omega \ \beta_{H_{i}^{0}}}{16384\ \pi^{3}  \  c_W^{2} v_{W}^{6}}     \nonumber  \\
&&  {\it Re}  \  \chi^{(1)}(\hat{s}) \ \chi^{(2)}(\hat{s})  \sum_{q=u,d}   \frac{\left (m_{u} \frac{v_{\eta}}{v_{\rho}}- m_{d} \frac{v_{\rho}}{v_{\eta}} \right )}{m_{q}}   I_q  \ \sum_{q=u,d}  I_q^{*}           ,
\end{eqnarray}

\begin{eqnarray} 
\left (\frac{d \hat{\sigma}}{d\cos \theta} \right )_{pp \rightarrow Z H_{2}^{0}}^{H_{1}^{0}} & = &   \frac{g^{2} \ \alpha_s^2 \  v_{\eta}^{2} \ v_{\rho}^{2} \ \hat{s} \ \Omega \ \beta_{H_{i}^{0}}}{8192 \ \pi^{3}  v_{W}^{6} \  c_W^{2}} \ |\chi^{(1)}(\hat{s})|^{2}  \left|
\sum_{q=u,d}   2 \delta_q + \delta_q (4 \ \delta_q-1)  I_q  \right|^2   + \nonumber   \\
&& + \frac{g^{2} \ \alpha_s^2 \ v_{\eta}^{2} \ v_{\rho}^{2}  \ \Omega \ \beta_{H_{i}^{0}}}{4096 \ \pi^{3} \ \hat{s} \  c_W^{2}  v_{W}^{6}}   \ |\chi^{(1)}(\hat{s})|^{2} \left|\sum_{q=u,d} m_{q}^{2}  I_q  \right|^2     ,
\end{eqnarray}

\begin{eqnarray} 
\left (\frac{d \hat{\sigma}}{d\cos \theta} \right )_{pp \rightarrow Z H_{2}^{0}}^{H_{2}^{0}} & = &  \frac{g^{2} \ \alpha_s^2  \ (v_{\rho}^{2} - v_{\eta}^{2} )^{2} \ \hat{s} \ \Omega \ \beta_{H_{i}^{0}}}{131072 \pi^{3}  v_{W}^{6}  \  c_W^{2}}  \ |\chi^{(2)}(\hat{s})|^{2}  \left|
\sum_{q=u,d} \frac{\left(m_{u} \frac{v_{\eta}}{v_{\rho}}- m_{d} \frac{v_{\rho}}{v_{\eta}} \right)}{m_{q}} \left[ 2 \delta_q + \delta_q (4 \ \delta_q-1)  I_q  \right] \right|^2    \nonumber   \\
&& + \frac{g^{2} \ \alpha_s^2  \ (v_{\rho}^{2}-  v_{\eta}^{2} )^{2}  \ \Omega \  \beta_{H_{i}^{0}}}{65536 \ \pi^{3} \hat{s} \  v_{W}^{6} \ c_W^{2}} \ |\chi^{(2)}(\hat{s})|^{2} \left|\sum_{q=u,d}  m_{q} \left (m_{u} \frac{v_{\eta}}{v_{\rho}}- m_{d} \frac{v_{\rho}}{v_{\eta}} \right ) I_q  \right|^2           ,
\end{eqnarray}

\begin{eqnarray} 
\left (\frac{d \hat{\sigma}}{d\cos \theta} \right )_{pp \rightarrow Z H_{2}^{0}}^{Z^{\prime}-H_{1}^{0}} & = & - \frac{g^{4} \ \alpha_s^2   \ \Lambda_{ZZ'H_{2}^{0}} \ v_{\eta} v_{\rho} \ \Pi \ \beta_{H_{i}^{0}}}{1024 \ \pi^{3} \  \hat{s}^{2} \  c_W^{2} v_{W}^{3}} {\it Re} \left[ \frac{\chi^{(1)}(\hat{s})}{(s- m_{Z'}^{2}+ im_{Z'} \Gamma_{Z'})}    \right.  \nonumber  \\
&& \left. \sum_{q=u,d}  m_{q}^{2} \ T_3^q \left( 1 + 2 \delta_q I_q  \right) \ \ \sum_{q=u,d}  I_q^{*} \right]         ,
\end{eqnarray}

\begin{eqnarray} 
\left (\frac{d \hat{\sigma}}{d\cos \theta} \right )_{pp \rightarrow Z H_{2}^{0}}^{Z^{\prime}-H_{2}^{0}} & = & - \frac{g^{4} \ \alpha_s^2   \ \Lambda_{ZZ'H_{2}^{0}} \ (v_{\rho}^{2}- v_{\eta}^{2}) \ \Pi \ \beta_{H_{i}^{0}}}{4096 \ \pi^{3} \  \hat{s}^{2} \  c_W^{2} v_{W}^{3}} {\it Re} \left[ \frac{\chi^{(1)}(\hat{s})}{(s- m_{Z'}^{2}+ im_{Z'} \Gamma_{Z'})}    \right.  \nonumber  \\
&& \left. \sum_{q=u,d}   m_{q} \left (m_{u} \frac{v_{\eta}}{v_{\rho}}- m_{d} \frac{v_{\rho}}{v_{\eta}} \right ) \ T_3^q \left( 1 + 2 \delta_q I_q  \right) \ \ \sum_{q=u,d}  I_q^{*} \right]              ,
\end{eqnarray}
where in Eq. (20) are considered the contribution of $Z^{\prime}$ boson, in Eq. (21) the contribution of interference of $H_1^0$ and $H_2^0$, in Eq. (22) and Eq. (23) the contribution of $H_1^0$ and  $H_2^0$, in Eq. (24) the contribution of interference of $Z^{\prime}$ and $H_1^0$ and in Eq. (25) the contribution of interference of $Z^{\prime}$ and $H_2^0$, all these contributions are to produce the $Z H_2^0$. The sum runs over all generations, $T_3^q$ is the quark weak isospin  [$T_{3}^{u(d)} = +(-)1/2$], ${\it Re}$ stands for the real part of the expression.  The loop function $I_{i} \equiv I(\delta_{i} = m_{i}^{2} /\hat{s})$, is defined by 

\begin{eqnarray*}
I_i \equiv I_i (\delta_i) = \int_0^1 \frac{dx}{x} 
\ln \left[1 - \frac{(1-x)x}{\delta_i} \right] =   
\left \{ \begin{array}{l}
- 2 \left[ \sin^{-1}\left( \frac{1}{2 \sqrt{\delta_{i}}} \right)\right]^2
\; , \;\;\; \delta_i > \frac{1}{4} \\ \nonumber
\frac{1}{2} \ln^2 \left(\frac{r_+}{r_-}\right) - \frac{\pi^2}{2}  + i\pi 
\ln\left(\frac{r_+}{r_-}  \right)  \; , \;\;\; \delta_i < \frac{1}{4} ,
\end{array}
\right.
\label{ii}
\end{eqnarray*}
with, $r_\pm = 1 \pm (1 - 4 \delta_i)^{1/2}$ and $\delta_i = 
m_i^2/\hat{s}$. Here, $i = q$ stands for the particle (quark ) running in the loop.

We have also defined $\Delta$, $\Omega$ and $\Pi$ which are equal to:

\[
\Delta = 4  \hat{s} - \frac{\hat{u}^{2}}{2  m_{Z}^{2}}  + \frac{\hat{t} \ \hat{u}}{m_{Z}^{2}} - \frac{\hat{t}^{2}}{2  m_{Z}^{2}}
\]

\[
\Omega = \frac{\hat{s}^{2}}{4 m_{Z}^{2}} - \frac{\hat{s}}{2} + \frac{m_{Z}^{2}}{4}- \frac{\hat{s} \ m_{H_{1}^{0}}^{2}}{2 m_{Z}^{2}} - \frac{m_{H_{1}^{0}}^{2}}{2} + \frac{m_{H_{1}^{0}}^{4}}{4 m_{Z}^{2}}
\]

\begin{eqnarray*}
\Pi & = & -\frac{\hat{s}^{4}}{8 \ m_{Z}^{4}} + \frac{\hat{s}^{3}}{4 \ m_{Z}^{2}} -\frac{\hat{s}^{2} \hat{u}}{8 \ m_{Z}^{2}} -\frac{\hat{s}^{2} \hat{t}}{8 \ m_{Z}^{2}} + \frac{\hat{s}^{2}}{8} + \frac{3 \ \hat{s} \hat{u}}{8} + \frac{3 \ \hat{s} \hat{t}}{8} - \frac{\hat{s} \ m_{Z}^{2}}{4}  +\frac{\hat{s}^{3} \ m_{H_{i}^{0}}^{2}}{4 \ m_{Z}^{4}}       \nonumber  \\ 
&& +\frac{\hat{s}^{2} \ m_{H_{i}^{0}}^{2}}{4 \ m_{Z}^{2}} +\frac{\hat{s} \ \hat{u} \ m_{H_{i}^{0}}^{2}}{8 \ m_{Z}^{2}}  +\frac{\hat{s} \ \hat{t} m_{H_{i}^{0}}^{2}}{8 \ m_{Z}^{2}} - \frac{3 \ \hat{s} \ m_{H_{i}^{0}}^{2}}{4} -\frac{\hat{s}^{2} \ m_{H_{i}^{0}}^{4}}{4 \ m_{Z}^{4}}   \nonumber  \ , 
\end{eqnarray*}

The total cross section for the process $pp \rightarrow gg \rightarrow  Z  H_{2}^{0}$ is related to the subprocess $gg \rightarrow Z  H_{2}^{0}$ total cross section $\hat{\sigma}$ through  
\begin{equation}   
\sigma = \int_{\tau_{min}}^{1}  \int_{\ln{\sqrt{\tau_{min}}}}^{-\ln{\sqrt{\tau_{min}}}} d\tau dy  G\left(\sqrt{\tau}e^y, Q^2\right) G\left(\sqrt{\tau}e^{-y}, Q^2\right)   \hat{\sigma}\left(\tau, s\right), 
\end{equation}
where $G\left(x,Q^2\right)$ is the gluon structure function and $\tau_{min}$ is given above.

\section{RESULTS AND CONCLUSIONS}   

In this work we have calculated contributions regarding to the Drell-Yan and gluon-gluon fusion in 3-3-1 model. We present the cross section for the process $pp \rightarrow Z H_{2}^{0}$ involving the Drell-Yan mechanism and the gluon-gluon fusion, to produce such Higgs bosons for the LHC. In all calculations we take for the parameters and the VEV the following representative values, \cite{TO96, cnt2} : $\lambda_{1} =0.3078$,  $\lambda_{2}=1.0$, $\lambda_{3}= -0.025$, $\lambda_{4}= 1.388$, $\lambda_{5}=-1.567$, $\lambda_{6}= 1.0$, $\lambda_{7} =-2.0$, $\lambda_{8}=-0.45$,  $v_{\eta}=195$ GeV, $\lambda_{9}=-0.90(-0.76,-0.71)$, correspond $v_{\chi}=1000(1500,2000)$ GeV, these parameters are used to estimate the values for the particles masses which are given in Table I, it is to notice that the value of $\lambda_{9}$ was chosen this way in order to guarantee the approximation $-f \simeq v_{\chi}$ \cite{TO96, cnt2}.

\begin{table}[h]
\caption{\label{tab1} Values for the particle masses used in this work.
All the values in this Table are given in GeV. Here, $m_{H^{\pm\pm}} =
500$ GeV and $m_T = 2v_\chi$.}
\begin{ruledtabular}
\begin{tabular}{c|ccccccccccccccc}
$f$ & $v_{\chi}$, $m_{J_1}$ & $m_E$ & $m_M$  & $m_{H_3^0}$ & $m_{h^0}$ &
$m_{H_1^0}$ & $m_{H_2^0}$ & $m_{H^\pm_2}$ & $m_V$ & $m_U$ & $m_{Z^\prime}$
& $m_{J_{2, 3}}$ \\
\hline
-1008.3 & 1000 & 148.9 & 875       & 2000   & 1454.6   & 126     & 1017.2
 & 183      & 467.5    & 464 & 1707.6 &1410 \\
-1499.7 & 1500 & 223.3 & 1312.5  & 474.34 & 2164.32 & 125.12 & 1525.8   &
387.23 &  694.12 & 691.76 & 2561.3 & 2115 \\
-1993.0 & 2000 & 297.8 & 1750     & 632.45 & 2877.07 & 125.12 & 2034.37 &
519.39 & 922.12  & 920.35 & 3415.12 & 2820 \\
\end{tabular}
\end{ruledtabular}
\end{table}



\subsection{The Higgs $H_{2}^{0}$}

The Higgs $H_{2}^{0}$ in 3-3-1 model is not coupled to a pair of standard bosons, it couples to quarks, leptons, Z Z', Z'Z' gauge bosons,  $H_{1}^{-} H_{1}^{+}$, $H_{2}^{-} H_{2}^{+}$, $h^{0}   h^{0}$, $H_{1}^{0}  H_{3}^{0}$ higgs bosons, $V^{-}V^{+}$ charged bosons, $U^{--} U^{++}$ double charged bosons, $H_{1}^{0} Z$, $H_{1}^{0} Z'$ bosons and $H^{--} H^{++}$ double charged Higgs bosons \cite{cieton2}. The Higgs $H_{2}^{0}$ can be much heavier than $ 1017.2$ GeV for $v_\chi = 1000 \ $GeV, $1525.8$ GeV for $v_\chi = 1500 \ $GeV, and $2034.37$ GeV for $v_\chi = 2000$ \ GeV, so the Higgs $H_2^{0}$ is a hevy particle. The coupling of the $H_{2}^{0}$ with $H_{1}^{0}$ contributes to the enhancement of the total cross section $\it{via}$ the Drell-Yan and gluon-gluon fusion. \par  

In Fig. 3 and 4, we show the cross section $pp \rightarrow Z H_{2}^{0}$ for Drell-Yan and gluon-gluon fusion, these processes will be studied for $\sqrt{s}=14$ TeV and for the vacuum expectation value $v_{\chi}=1000$ GeV, $v_{\chi}=1500$ GeV, and $v_{\chi}=2000$ GeV. Considering that the expected integrated luminosity for the LHC collider that will be reach is of order of $300 $ fb$^{-1}$ then the statistics for $v_{\chi}=1000$ GeV gives a total of   $\simeq 2.7 \times 10^5(1.4 \times 10^5)$ events per year for Drell-Yan and $\simeq 9.3 \times 10^4(7.5 \times 10^4)$ events per year for gluon-gluon fusion, if we take the mass of the Higgs boson $m_{H_{2}^{0}}= 1100(1300)$ GeV ($\Gamma_{H_{2}^{0}} = 878.25, 1091.33$ GeV) and correspond to 14 TeV for the LHC, respectively. These values are in accord with the Table I. It must be noticed that must take care with large Higgs masses, as the width approaches the value of the mass itself for a very heavy Higgs and one looses the concept of resonance. \par 

To obtain event rates we multiply the production cross sections by the respective branching ratios. Considering that the signal for  $H_{2}^{0}Z$ production for  $m_{H_{2}^{0}}= 1100(1300)$ GeV and $v_{\chi}=1000$ GeV will be $H_{2}^{0} Z \rightarrow  Z H_{1}^{0} Z$, and taking into account that the branching ratios for these particles would be $BR(H^{0}_{2} \to Z H_{1}^{0}) = 39.5 (43.4)  \ \% $,  see Fig. 5, and $BR(Z \to  b \bar{b}) = 15.2 \ \% $, and that the particles $H_{1}^{0}$ decay into  $W^{+} W^{-}$, and taking into account that the branching ratios for these  particles would be $BR(H^{0}_{1} \to W^{+} W^{-}) = 23.1 \ \% $ followed by leptonic decay of the boson $W^{+}$ into $\ell^{+} \nu$ and $W^{-}$ into $\ell^{-} \bar{\nu}$ whose branching ratios for these particles would be  $BR(W  \to \ell \nu) = 10.8 \ \%$, then we would have approximately  $ \simeq 7 (4))$  events per year for Drell-Yan and $\simeq 2(2)$ for gluon-gluon fusion for the signal $b\bar{b} b \bar{b} \ell^{+} \ell^{-} X$. \par 

The statistics for $v_{\chi}=1500$ gives a total of   $\simeq 7.1 \times 10^4( 4.5 \times 10^4)$ events per year for Drell-Yan and $\simeq 2.8 \times 10^3(2.5 \times 10^3)$ events per year for gluon-gluon fusion, if we take the mass of the Higgs boson $m_{H_{2}^{0}}= 1600(1800)$ GeV, respectively. These values are in accord with the Table I. Taking into account the same signal as above, that is $H_{2}^{0} Z \rightarrow  Z H_{1}^{0} Z$, and taking into account that the branching ratios for these particles would be $BR(H^{0}_{2} \to Z H_{1}^{0}) = 44.2 (45.9)  \ \% $, see Fig. 6, $BR(Z \to  b \bar{b}) = 15.2 \ \% $, $BR(H^{0}_{1} \to W^{+} W^{-}) = 23.1 \ \% $, $BR(W  \to \ell \nu) = 10.8 \ \%$, we would have approximately  $ \simeq 2 (1))$  events per year for Drell-Yan and $\simeq 0(0)$ for gluon-gluon fusion for the signal $b\bar{b} b \bar{b} \ell^{+} \ell^{-} X$. \par 

With respect to vacuum  expectation  value $v_{\chi}=2000$ GeV, for the masses of $m_{H_{2}^{0}}= 2100(2300)$ it  will give a total of  $\simeq 2.4 \times 10^4(1.6 \times 10^4 )$  events per year to produce $H_{2}^{0}$ for Drell-Yan and in respect to gluon-gluon fusion we will have  $ \simeq 119 (107 )$ events per year to produce the same particles. Taking into account the same signal as above, that is $b\bar{b} b \bar{b} \ell^{+} \ell^{-} X$ and considering  that the branching ratios for $H_{2}^{0}$ would be $BR(H^{0}_{2} \to Z H_{1}^{0}) = 46.4 (47.3)  \ \% $,  see Fig. 7, , we will have approximately $ \simeq 0(0)$ events per year for Drell-Yan and $\simeq 0 (0)$ for gluon-gluon fusion. \par

The main background to this signal is $t \bar{t} Z \rightarrow b\bar{b} b \bar{b} \ell^{+} \ell^{-} X$, which cross section at LO is $\simeq 1$ pb for $\sqrt{s}=14$ TeV. Considering that the $t \bar{t}$ particles decay into $b \bar{b} W^{+} W^{-}$, whose branching ratios for these particles would be $BR(t \rightarrow bW) = 99.8 \%$ followed by leptonic decay of the boson W, that is $BR(W \rightarrow \ell \nu) = 10.8 \%$ and $BR(Z \rightarrow b \bar{b}) = 15.2 \%$ then we would have approximately a total of $ \simeq 530$ events for the background and $\simeq 9(6)$ events for the signal for $m_{H_{2}^{0}}= 1100(1300)$ GeV and $v_{\chi}=1000$; on the other hand, for $v_{\chi}=1500$ and $v_{\chi}=2000$ the number of events for the signal is insignificant. \par 

Therefore we have that the statistical significance is $\simeq 0.39(0.26) \sigma$ for $m_{H_{2}^{0}}= 1100(1300)$ GeV, that is a low probability to detect
the signals. The improvement will be significant if we consider a luminosity $\simeq 10$ times higher than original LHC design, that is what we are awaiting to happen for 2025, then we will have $\simeq 90(60)$ events for the signals for $m_{H_{2}^{0}}=1100(1300)$ GeV and $v_{\chi}=1000$, which corresponds to have a $\simeq 3.9(2.6) \sigma$, then we have an evidence for $\simeq 3.9 \sigma$ discovery in the $b\bar{b} b \bar{b} \ell^{+} \ell^{-} X$ final state. On the other hand, for $v_{\chi}=1500$ we have $\simeq 20(10)$ events for $m_{H_{2}^{0}}= 1600(1800)$ GeV and which corresponds to $\simeq 0.89(0.44) \sigma$, for this last scenario the signal is too small to be observed even with 3000 $fb^{-1}$. \par 

To extract the signal from the background we must select the $b \bar{b}$ channel using the techniques of b-flavour identification, thus reducing the huge QCD backgrounds of quark and gluon jets. Later, the Z that comes together with the $H_{2}^{0}$ and the other Z that comes from the decay of $H_{2}^{0}$ would appear as a peak in the invariant mass distribution of b-quark pairs. The charged lepton track from the $W$ decay and the cut on the missing transverse momentum ${p\!\!\slash}_{T} >$ 20 GeV allows for a very strong reduction of the backgrounds. \par 

The $H_{2}^{0} Z$ will also decay into  $t \bar{t} \ \ell^{+} \ell^{-}$, and consider that the branching ratios for these particles would be  $BR(H^{0}_{2} \to  t \bar{t}) = 5.1 (4.1)  \ \% $,  see Fig. 5 and $BR(Z \to \ell^{+} \ell^{-}) = 3.4 \ \% $ for the mass of the Higgs boson $m_{H_{2}^{0}}= 1100(1300)$ GeV and $v_{\chi}=1000$ GeV and that the particles $t \bar{t}$ decay into $ b \bar{b} W^{+} W^{-}$, whose branching ratios for these particles would be  $BR(t \to b W) = 99.8 \ \% $, followed by leptonic decay of the boson W, that is  $BR(W \to e \nu) = 10.75 \ \% $, then  we would have approximately $\simeq 5 (2)$ events per year for Drell-Yan. Regarding to gluon-gluon fusion we will have  $ \simeq 2(1)$ events per year to produce the same particles. Considering the vacuum  expectation value $v_{\chi}=1500$ GeV and the branching ratios $BR(H_{2}^{0} \rightarrow t \bar{t}) = 2.8 (2.3) \ \% $, see Fig. 6 and taking the same parameters and branching ratios for the same particles given above, then we would have for  $m_{H_{2}^{0}}= 1600(1800)$ a total of  $ \simeq 1(1)$ events of $H_{2}^{0}$ produced per year for Drell-Yan and in respect to gluon-gluon fusion the number of events per year for the signal will be $ \simeq 0(0)$. \par 

Taking again the irreducible background $t \bar{t} Z  \rightarrow b \bar{b} e^{+} e^{-} \ell^{+} \ell^{-} X$, and using CompHep \cite{pukhov} we have that a cross section at LO is $\simeq 1$ pb, which gives $ \simeq 117$ events. So we will have a total of $\simeq  6(3)$ events per year for the signal for $m_{H_{2}^{0}}= 1100(1300)$ GeV and $v_{\chi}=1000$ and for $v_{\chi}=1500$ and $v_{\chi}=2000$, the number of events is insignificant. \par 

Then we have that the statistical significance is $\simeq 0.55(0.28) \sigma$ for $m_{H_{2}^{0}}= 1100(1300)$ GeV and $v_{\chi}=1000$ GeV. For this scenario the signal significance is smaller than 1$\sigma$ and discovery can not be accomplished unless the luminosity will be improved. So, if we enhance the integrated luminosity up to 3000 $fb^{-1}$, then we will have $\simeq 60(30)$ events for the signals for $m_{H_{2}^{0}}=1100(1300)$ GeV and $v_{\chi}=1000$, which corresponds to have a $\simeq 5.5(2.8) \sigma$ discovery in the $b \bar{b} e^{+} e^{-} \ell^{+} \ell^{-} X$ final state and for $v_{\chi}=1500(2000)$ the signal will be not visible in this channel. We impose the following cuts to improve the statistical significance of a signal, i. e. we isolate a hard lepton from the $W$ decay with $p_{T}^{\ell}>$ 20 GeV, put the cut on the missing transverse momentum ${p\!\!\slash}_{T} >$ 20 GeV and apply the Z window cut $|m_{\ell^{+} \ell^{-}} - m_{Z}| >$ 10 GeV, which removes events where the leptons come from Z decay \cite{cheng}. However, all this scenarios can only be cleared by a careful Monte Carlo work to determine the size of the signal and background. \par   

In summary, we showed in this work that in the context of the 3-3-1 model the signatures for neutral Higgs boson $H_2^0$ can be significant in LHC collider if we take $v_{\chi}=1000$, $m_{H_{2}^{0}}=1100(1300)$ GeV, $\sqrt{s}=14$ TeV and a luminosity of 3000 $fb^{-1}$. In other scenarios, the signal will be too small to be observed even with 3000 $fb^{-1}$. Our study indicates the possibility of obtaining a signal of this new particle in the channel $t \bar{t} Z  \rightarrow b \bar{b} e^{+} e^{-} \ell^{+} \ell^{-} X$. If this model is realizable in the nature, certainly new particles will appear such as $H_{2}^{0}, Z^{\prime}$ in the context of this study. \par

\acknowledgments
{MDT is grateful to the Instituto de F\'\i sica Te\'orica of
the UNESP for hospitality, the Brazilian agencies CNPq for a research grant, and FAPESP for financial support (Processo No. 2009/02272-2).}

\newpage
\newpage

\begin{center}
FIGURE CAPTIONS
\end{center}

  
{\bf Figure 1}: Feynman diagrams for production of neutral Higgs {\it via} Drell-Yan process.
 
{\bf Figure 2}: Feynman diagrams for production of neutral Higgs {\it via} gluon-gluon fusion.  

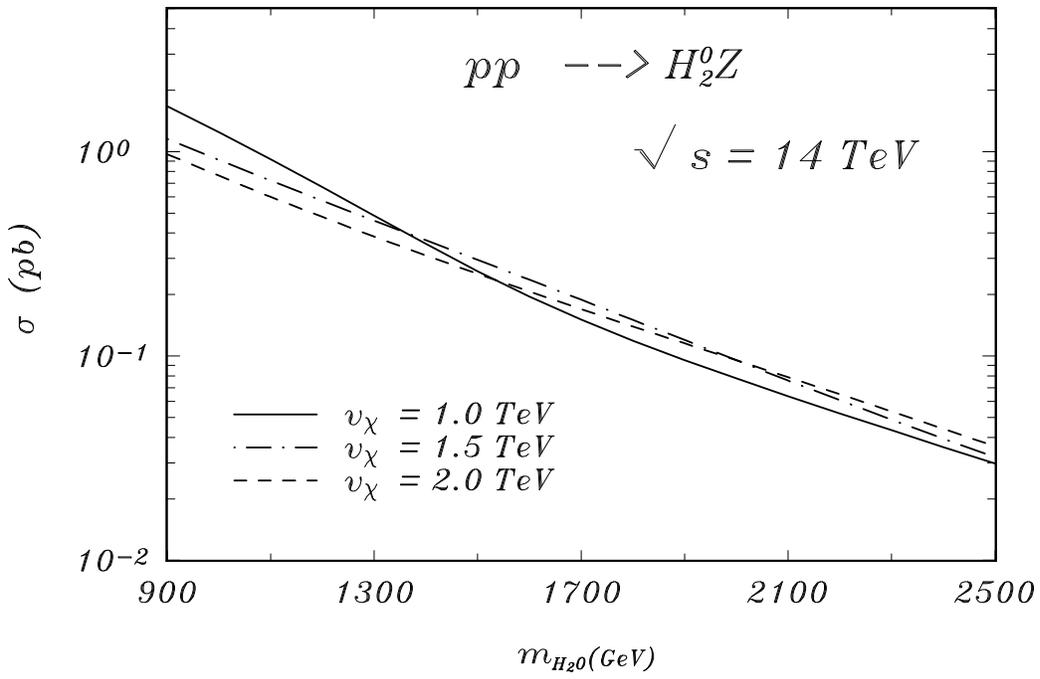
{\bf figure 3}:Total cross section for the process $p \ p \to
Z H_{2}^{0}$ as a function of $m_{H_{2}^{0}}$ at $\sqrt{s} = 14$ TeV $\it{via}$ Drell-Yan. Solid line for $v_{\chi}=1.0$ TeV, dot-dash line for $v_{\chi}= 1.5$ TeV and short dash line for $v_{\chi}= 2.0$ TeV;

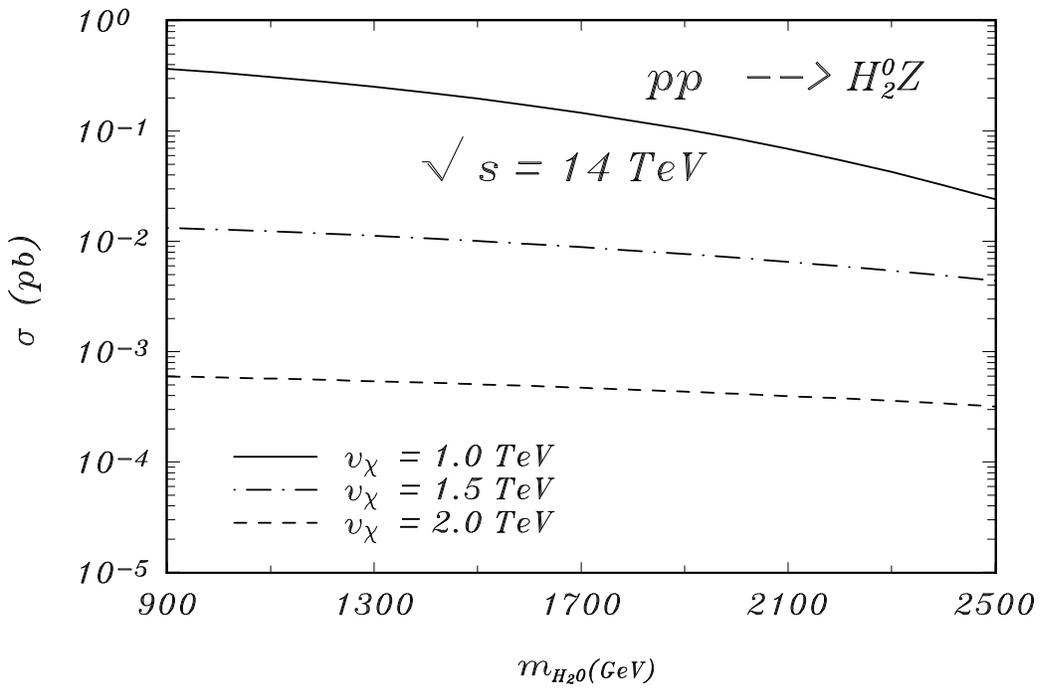
{\bf figure 4}:Total cross section for the process $p \ p \to
Z H_{2}^{0}$ as a function of $m_{H_{2}^{0}}$ at $\sqrt{s} = 14$ TeV $\it{via}$ gluon-gluon fusion. Solid line for $v_{\chi}=1.0$ TeV, dot-dash line for $v_{\chi}= 1.5$ TeV and short dash line for $v_{\chi}= 2.0$ TeV ;

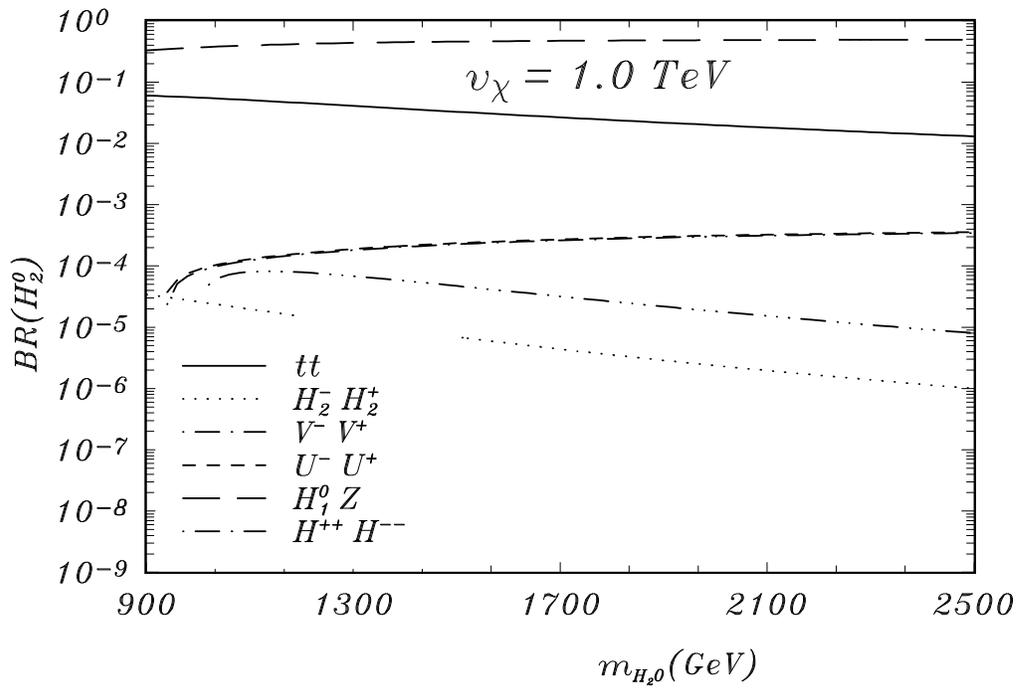
{\bf figure 5}: BRs for the $H_{2}^{0}$ decays as functions of $m_{H_{2}^{0}}$ for $v_{\chi} = 1.0$ TeV.

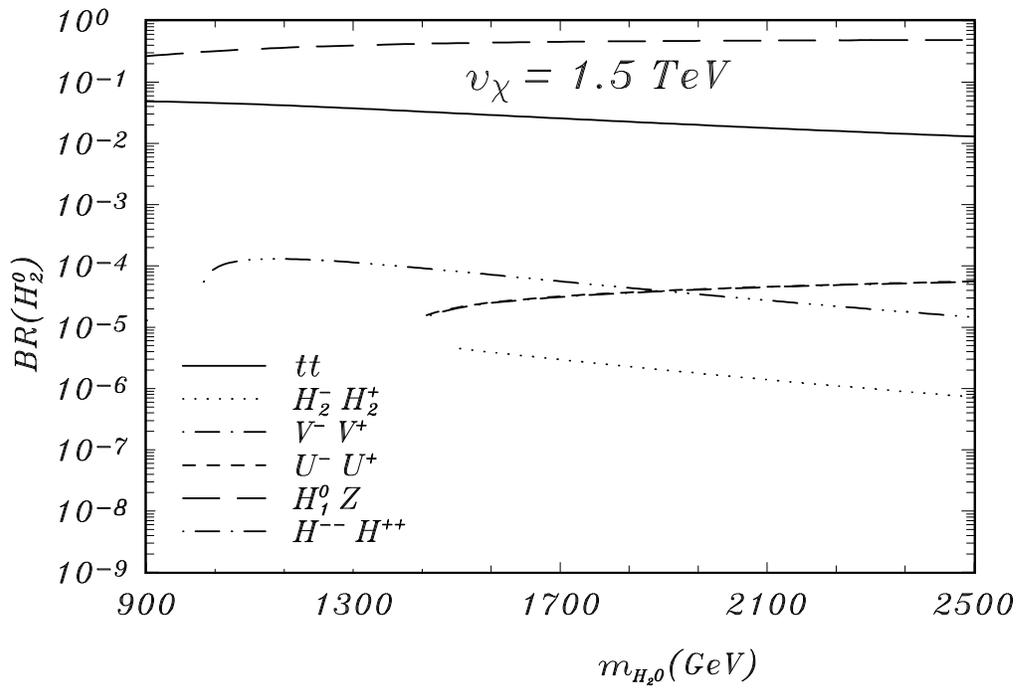
{\bf figure 6}: BRs for the $H_{2}^{0}$ decays as functions of $m_{H_{2}^{0}}$ for $v_\chi = 1.5$ TeV.

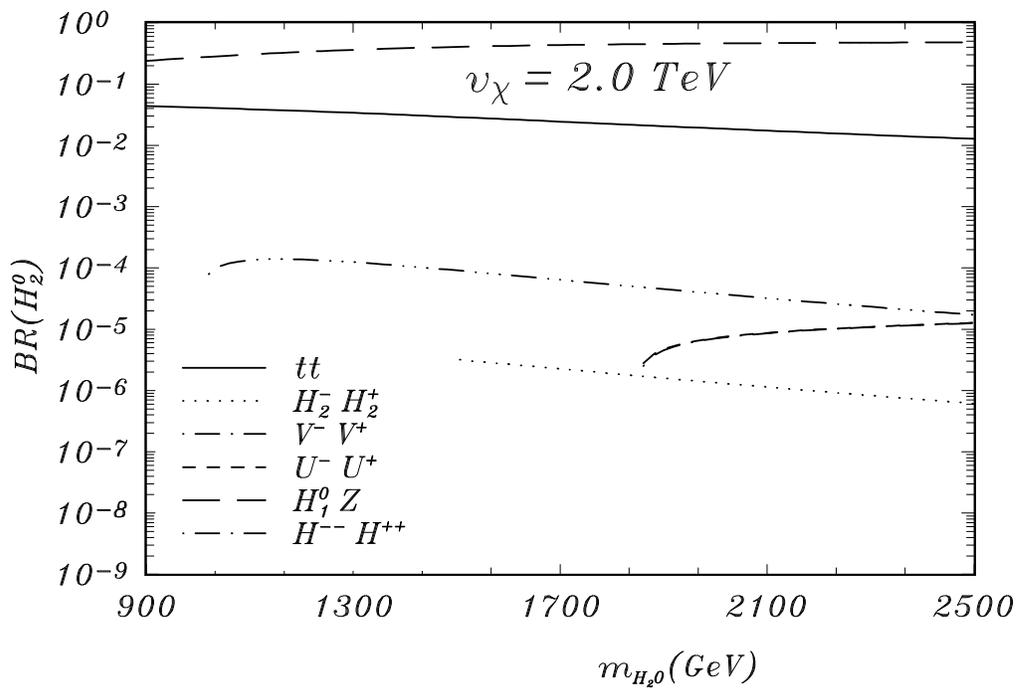
{\bf figure 7}: BRs for the $H_{2}^{0}$ decays as functions of $m_{H_{2}^{0}}$ for $v_\chi = 2.0$ TeV.

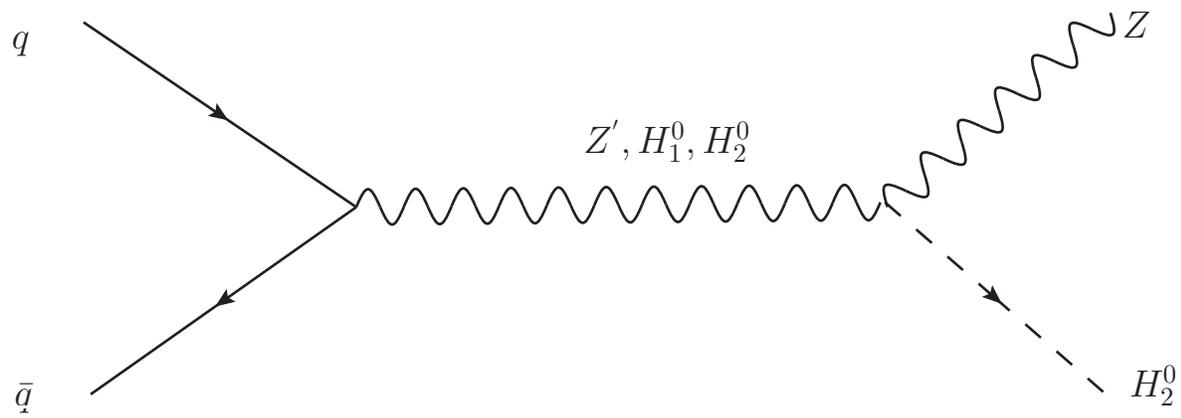
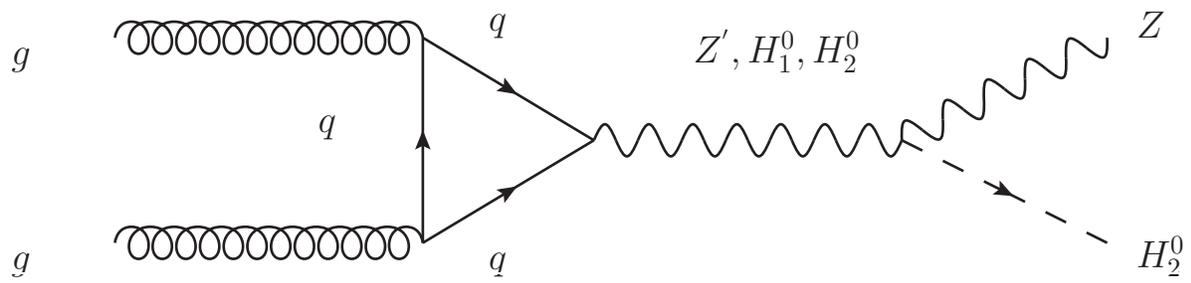
\end{document}